\documentclass[a4paper,11pt]{article}
\pdfoutput=1 % if your are submitting a pdflatex (i.e. if you have
\usepackage{jcappub} % for details on the use of the package, please
\usepackage{graphicx}
\usepackage[T1]{fontenc} % if needed
\usepackage{xcolor}
\usepackage{subcaption}
\usepackage{mwe}

\title{Cosmic Inflation: The Most Powerful Microscope in the Universe\footnote{Essay written for the Gravity Research Foundation 2018 Awards for Essays on Gravitation.\\Submitted March 30 2018}}

\author[a]{Andreas Albrecht,}
\author[a,c]{Nadia Bolis,}
\author[b]{R.~Holman\footnote{Corresponding Author}}
\affiliation[a]{Center for Quantum Mathematics and Physics, and Department of Physics, University of California at Davis, One Shields Ave, Davis, CA 95616, USA}
\affiliation[b]{College of Computational Sciences, Minerva Schools at KGI, 1145 Market Street, San Francisco, CA 94103, USA}
\affiliation[c]{Central European Institute for Cosmology and Fundamental Physics (CEICO), \\Fyzik\'{a}ln\'{i} \'{u}stav Akademie v\v{e}d \v{C}R, Na Slovance 1999/2, CZ-182 21 Praha 8, Czech Republic}

\emailAdd{nbolis@ucdavis.edu}
\emailAdd{ajalbrecht@ucdavis.edu}
\emailAdd{rholman@minerva.kgi.edu}

\abstract{How well can we constrain the initial quantum state of metric perturbations sourced during inflation? We exhibit an interesting new class of quantum states that entangle the scalar metric perturbations $\zeta$ with other fields such as scalars as well as the tensor metric perturbations $h_{i j}$. These states are theoretically consistent, for inflation that lasts close to its minimum number of e-folds. They give distinguishable signatures in the power spectrum and may be able to explain some long-standing anomalies in the CMB power spectrum. We advocate using a generalized effective theory of quantum states (of which our work is an example) that, using inflation as a powerful microscope, could provide deep insights into the quantum state of matter on the smallest scales.}

\begin{document}
\maketitle
\flushbottom

\section{A Manifesto}

One of the extraordinary aspects of cosmic inflation is that it takes physical features on the very smallest scales and expands them into observable features in the cosmos.  Furthermore, when one takes the most conventional view of these small scale features (namely that they describe the vacuum state of an effective field theory (EFT)) the observational predictions match the data to a remarkable degree. The fact that our ideas about the EFT vacuum are linked with the independent successes of quantum EFTs in describing elementary particles suggests we are achieving a deep picture unifying nearly all physical scales.

In addition, one can develop lofty principles such as ``naturalness'' to validate the use of a given vacuum state, which would seem to lend further credence to this picture. But here we articulate  a different point of view:  The machinery of inflation is in fact the most powerful microscope in the Universe, providing an opportunity to test a variety of alternatives to the standard vacuum picture by mapping these alternatives onto potentially observable cosmic signatures.  If found, such signatures would be thrilling breakthrough discoveries.  In the case of null results we would be giving these principles additional observational backing, so they would not need to stand on loftiness alone.  We also note that various modern ideas in theoretical physics (such as the black hole firewall problem, the entanglement/geometry correspondence or simply the notion that spacetime itself is emergent from some very different physical picture) are bringing  the EFT vacuum under new scrutiny.

In our picture the standard vacuum is just one of many possible initial states for inflation. It's worth recounting the standard dogma concerning these initial states. One solves the equation of motion for the inflaton fluctuation modes $u_k(\eta)$, where $k$ is the comoving wavenumber and $\eta\in (-\infty, 0]$ is conformal time. The solution is parametrized by two parameters. One of these is fixed by the canonical commutation relations between the field and its canonically conjugate momentum, leaving one yet to be determined. It is this one that tells the story. 

IF we can take the limit $k\eta\rightarrow -\infty$, we are to choose the undetermined constant such that in this limit, the mode function reduces to the mode function for the flat space vacuum:
$$
u_k(\eta)\rightarrow \frac{1}{\sqrt{2 k}} e^{-ik\eta}\ {\rm as}\ k\eta\rightarrow -\infty
$$
These modes define the Bunch-Davies (BD) state\cite{Bunch:1978yq} of a quantum field evolving in a de Sitter-like spacetime. The BD vacuum has a variety of useful properties. It is a state of infinite adiabatic order so it can truly be considered to be devoid of particles. This ensures that there is no backreaction of particles to the stress energy tensor that might {\em prevent} the onset of inflation. 

The Bunch-Davies state seems like the obvious choice of initial state and it has several advantages. For example, at short distance scales we can make use of the Principle of Equivalence to argue that fields should behave as if they see flat space and thus their quantum fluctuations should find themselves in the state that is the nearest approximation to the flat space vacuum; the BD state. It has also been argued that if inflation lasts for longer than the $60-70$ e-folds (``long'' inflation) the BD state is a strong attractor in the space of states \cite{Anderson:2005hi,Kaloper:2018zgi}.

We now come to our manifesto.  What do we actually know? Current cosmological data\cite{Ade:2015xua} is consistent with the predictions of the concordance $\Lambda$CDM model, with an initial power spectrum of metric perturbations of the form arising from an inflationary phase lasting $60-70$ e-folds.  There is {\em no} empirical evidence that requires long inflation. Rather, the desire for long inflation  appears to be driven more by ideas of ``naturalness'': short or just-so inflation seems to require fine tuning (although some holographic arguments actually prefer short inflation\cite{Phillips:2014yma}). 

In order to truly validate (or refute) the BD vacuum as the initial state of the metric perturbations  in inflation, it is important to find a way to observationally distinguish and constrain alternate scenarios. 
If we are solely motivated by the physical evidence at hand, and if we work within the inflationary universe paradigm,  it would be unreasonable to just assume, without further evidence, that inflation continued past the observed $60-70$ e-folds.  In this context, it makes sense to consider short inflation models, revisit the choice of the Bunch-Davies state as the initial state, and let the data dictate the correct choice(s).

\section{Constraints on the Choice of Initial States}
\label{sec:constraints}

In general non-BD states have a non-zero particle production that might derail inflation entirely: this is called the back reaction problem.  We need to demand that the energy density coming from these states, $\rho = \langle T^0_0\rangle$ must be smaller than $M^2_{\rm Pl}H^2$, where $H$ is the Hubble parameter during inflation. This is only possible for states defined from some finite initial (conformal) time $\eta_0$, which certainly fits with our short inflation approach.

\section{Entangled Initial States}
\label{sec:entangled}

The space of possible consistent initial states is surely quite large, especially once the constraint of long inflation is relaxed. In particular, given the existence of fields other than the metric and the inflaton in Nature, there is no reason the initial state needs to treat the metric perturbations separately from the fluctuations in the other fields. This led us to the speculation that perhaps some early universe effect {\em entangled} the metric  and field perturbations in a non-trivial way. 

The simplest realization of this would be the metric perturbation $\zeta$ entangled with another scalar field $\phi$\cite{Albrecht:2014aga} which might be the Higgs field or one of the plethora of scalars that can appear in extensions of the standard model (other realizations are considered in \cite{Bolis:2016vas,Collins:2016ahj}. 

The Schr\"odinger field theory formalism is especially well suited to the analysis of entangled states.  In this formalism a {\em wavefunctional} $\Psi\left[\zeta,\phi; \eta\right]$, describes the probability amplitude for finding field configurations $\zeta(\vec{x}),\ \phi(\vec{x})$ at conformal time $\eta$. We can write a Gaussian entangled state as:
\begin{eqnarray}
\label{eq:wavefunctional}
&& \Psi\left[\zeta,\phi; \eta\right] =\prod_{\vec{k}} \psi_{\vec{k}}\left(\zeta_{\vec{k}},\phi_{\vec{k}}; \eta\right)\nonumber\\
&& \psi_{\vec{k}}\left(\zeta_{\vec{k}},\phi_{\vec{k}}; \eta\right)=N_k(\eta)\exp\left(-\frac{1}{2} A_k(\eta)\zeta_{\vec{k}} \zeta_{-\vec{k}} -\frac{1}{2} B_k(\eta)\phi_{\vec{k}} \phi_{-\vec{k}}-\frac{1}{2} C_k(\eta)\left(\zeta_{\vec{k}} \phi_{-\vec{k}}+\phi_{\vec{k}} \zeta_{-\vec{k}}\right)\right),\nonumber\\
\end{eqnarray}
where $\zeta_{\vec{k}},\ \phi_{\vec{k}}$ are the Fourier modes of $\zeta(\vec{x}),\ \phi(\vec{x})$ (recall the in the Schr\"odinger picture operators have no time dependence) and, for simplicity,  we assume spatially flat spatial sections for our geometry.   The kernels $A_k(\eta),\ B_k(\eta),\ C_k(\eta)$ are then obtained by solving the Schr\"odinger equation; for consistency we use the quadratic part of the Hamiltonian, which itself factorizes in terms of the spatial momenta:

\begin{equation}
\label{eq:Seqn}
i\partial_{\eta} \psi_{\vec{k}}\left(\zeta_{\vec{k}},\phi_{\vec{k}}; \eta\right)=(H^{(\zeta)}_{\vec{k}}+H^{(\phi)}_{\vec{k}}) \psi_{\vec{k}}\left(\zeta_{\vec{k}},\phi_{\vec{k}}; \eta\right),
\end{equation}
The Hamiltonians $H^{(\zeta)}_{\vec{k}}+H^{(\phi)}_{\vec{k}}$ are given by

\begin{eqnarray}
\label{eq:Hams}
H^{(\zeta)}_{\vec{k}}&=&\frac{\Pi_{\vec{k}} \Pi_{-\vec{k}}}{2 \alpha^2} + \frac{k^2 \alpha^2}{2} \zeta_{\vec{k}} \zeta_{-\vec{k}}\nonumber\\ 
H^{(\phi)}_{\vec{k}}&=& \frac{\pi_{\phi, \vec{k}} \pi_{\phi, -\vec{k}}}{2 a^2(\eta)} + \frac{1}{2} a^2(\eta)\left(k^2 + m_{\phi}^2 a^2(\eta)\right) \phi_{\vec{k}} \phi_{-\vec{k}},
\end{eqnarray}
with $\Pi_{\vec{k}},\ \pi_{\phi, \vec{k}} $ being the momenta conjugate to $\zeta_{-\vec{k}}\ ,\phi_{-\vec{k}}$ respectively, $a(\eta)$ is the scale factor in conformal time and $\alpha^2 = a^2(\eta) \epsilon M^2_{\rm Pl}$, where $\epsilon$ is the slow-roll parameter during inflation.

%\begin{figure}[!htbp]

%\centering % \begin{center}/\end{center} takes some additional vertical space
%\includegraphics[width=.6\textwidth]{Pkplot.png}
%\caption{The primordial power spectrum from entangled states for various values of $\lambda_k$ (entanglement strength). The primordial power spectrum displays oscillations in $k$ whose amplitudes increase with $\lambda_k$.}
 %\end{figure}

\begin{figure}[!htbp]
    \centering
    \begin{subfigure}[b]{0.75\textwidth}
        \includegraphics[width=\textwidth, height=6.5cm]{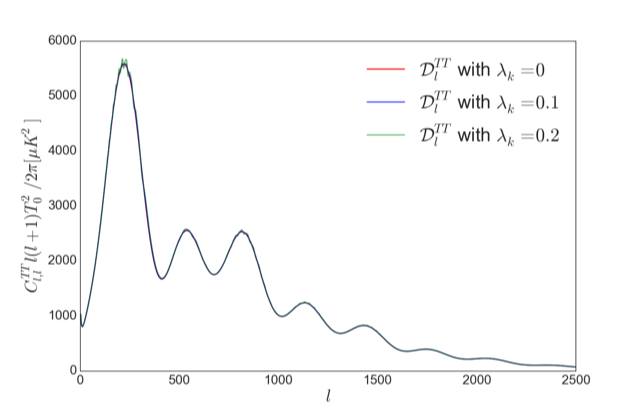}
        \caption{ }
    \end{subfigure}
    ~ %add desired spacing between images, e. g. ~, \quad, \qquad, \hfill etc. 
      
    \begin{subfigure}[b]{0.75\textwidth}
        \includegraphics[width=\textwidth, height=6.5cm]{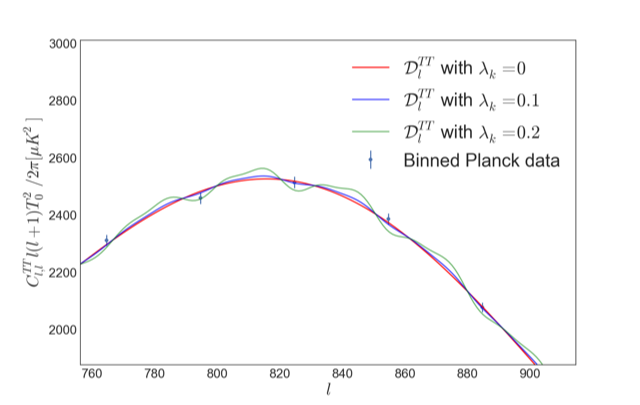}
    \caption{ }
    \end{subfigure}
    \caption{\textbf{(a)} The CMB power spectra from entangled states for various values of $\lambda_k$. We include $\lambda_k=0$ for comparison. \textbf{(b)} A zoomed in portion of the CMB power spectra in (a) with the binned Planck data.}
    \label{fig:powerspectrum}
\end{figure}

In \cite{Albrecht:2014aga} we show that the effects of the entanglement kernel $C_k(\eta)$ can be encoded in a time independent parameter $\lambda_k$. Figure~\ref{fig:powerspectrum} shows how the power spectrum of the CMB changes with $\lambda_k$. The current data can provide interesting bounds on $\lambda_k$, and may even contain a signal.

\section{Coda}
\label{sec:coda}

What's the take-home lesson from this analysis? If we stick to our pragmatic ``inflation as microscope'' approach, we should really be thinking of an effective theory of {\em states}, parametrized by various kernels as in the states above.  Using both theory inputs, such as symmetries and the backreaction constraint, together with the available data we can restrict the possible forms of the kernels. In doing so,  this framework (the EFT of states) can be exploited to constrain and quantify how close the initial state of the inflaton was to the BD vacuum. 
The entangled states displayed above fall within the scope of this framework. They can clearly be distinguished from the Bunch-Davies states in the resulting power spectra (and bi-spectra\cite{usbispectrum}), allowing us to constrain how much entanglement is allowed.

Ultimately, we view this program of exploring the space of states as a no-lose proposition. If we can find a state that explains phenomena, such as the CMB anomalies that the BD state cannot, that gives us an exciting new direction to explore. If we eventually have enough data to make states other than the BD one unviable, we achieve the grand goal of finding {\em the} initial quantum state of the inflaton. Surely this too would represent a magnificent triumph of the scientific method.

\acknowledgments

AA was supported by UC Davis. NB acknowledges funding from the European Research Council under the European Union's Seventh Framework Programme (FP7/2007-2013)/ERC Grant Agreement No. 617656 ``Theories and Models of the Dark Sector: Dark Matter, Dark Energy and Gravity.''  RH was supported by Minerva Schools at KGI.


\bibliographystyle{JHEP}
\bibliography{GRFEssay}

\end{document}